\documentclass[11pt,letterpaper]{article}
\pdfoutput=1
\usepackage{jcappub}

\usepackage{appendix}
\usepackage{array}
\newcolumntype{x}[1]{>{\centering\arraybackslash}p{#1}}

\usepackage{amsmath}
\usepackage{amsfonts}
\usepackage{amssymb}
\usepackage{eucal}               
\usepackage{mathrsfs}          
\usepackage{slashed}

\usepackage{epsfig}
\usepackage{graphicx}
\usepackage{dcolumn}
\usepackage{enumerate}

\usepackage{color}
\usepackage{ulem}
\hypersetup{colorlinks=false}

\definecolor{switte}{rgb}{0.558, 0.188, 0.878}

\newcommand{\eg}{e.g.~}
\newcommand{\ie}{i.e.~}
\newcommand{\Eq}[1]{Eq.~\eqref{#1}}
\newcommand{\Fig}[1]{Fig.~\ref{#1}}

\newcommand{\beq}{\begin{equation}}
\newcommand{\eeq}{\end{equation}}
\newcommand{\ud}{\text{d}}
\newcommand{\bol}[1]{\boldsymbol{#1}}
\newcommand{\ER}{E_\text{R}}

\newcommand{\vesc}{v_\text{esc}}
\newcommand{\vmin}{v_\text{min}}

\title{Gravitational Focusing and Substructure Effects on the Rate Modulation in Direct Dark Matter Searches}

\author{Eugenio Del Nobile,}
\author{Graciela B.~Gelmini,}
\author{Samuel J.~Witte}
\affiliation{Department of Physics and Astronomy, UCLA,\\
475 Portola Plaza, Los Angeles, CA 90095, USA}

\emailAdd{delnobile@physics.ucla.edu}
\emailAdd{gelmini@physics.ucla.edu}
\emailAdd{switte@physics.ucla.edu}

\abstract{We study how gravitational focusing (GF) of dark matter by the Sun affects the annual and biannual modulation of the expected signal in non-directional direct dark matter searches, in the presence of dark matter substructure in the local dark halo. We consider the Sagittarius stream and a possible dark disk, and show that GF suppresses some, but not all, of the distinguishing features that would characterize substructure of the dark halo were GF neglected.} 


\notoc

\begin{document}
\maketitle
\flushbottom

\newpage

\section{Introduction}
\label{sec:intro}

Astrophysical and cosmological observations indicate that dark matter (DM) is the dominant form of matter in the Universe. One of the most well-motivated candidates for DM is the weakly interacting massive particle (WIMP) \cite{GelminiTasiLecture2014}.
WIMPs are particles with weak-scale interaction cross sections, and with masses roughly between a few GeV and hundreds of TeV. They could be detected through their scattering with atomic nuclei using sensitive, low threshold detectors. Many such direct detection experiments are currently in operation, employing a variety of target nuclei and detection techniques, attempting to gain further insight into the exact nature of DM.

Due to Earth's rotation around the Sun, we expect the DM flux seen at Earth, and therefore the scattering rate at non-directional direct detection experiments, to be annually modulated. While the signal of WIMP scattering off of target nuclei can be faked, for example by scattering of neutrons emitted by radioactive processes in the vicinity of the detector, an annual modulation in the rate with the expected features is a much more difficult signature to be reproduced by spurious sources. Moreover, while the energy spectrum of DM events depends on the WIMP mass and interactions, a modulation in the rate will be present regardless of these details and it is therefore sometimes claimed to be a model-independent signature of DM.

In order to determine the modulation amplitude and its spectrum in energy measured by a particular experiment, a model for the DM halo must be assumed. The standard choice for the main virialized component is the Standard Halo Model (SHM), an isothermal sphere at rest with respect to the galaxy with an isotropic Maxwellian velocity distribution \cite{FreeseLisantiSavageReview}. Despite its simple form, the SHM is believed to capture the relevant characteristics of the dark halo and thus has been widely used in the literature. However, we expect the actual halo to deviate from this simple model. The local density and velocity distribution could actually be very different if Earth is within a DM clump (although this is unlikely \cite{Vogelsberger:2008qb}), stream, dark disk (DD), and/or tidal debris \cite{Green,GelminiGondolo,FreeseGondoloNewbergLewis,FreeseGondoloNewberg,SavageFreeseGondolo,FornengoScopel,BelliCerulliFornengoScopel,KuhlenLisantiSpergel, PurcellBullockKaplinghat,ReadLakeAgertzDebattista,BruchReadBaudisLake,ReadMayerBrooksGovernatoLake,FanKatzRandallReece, PurcellZentnerWang, Lisanti:2011as}. 

Both a stream and a DD are well motivated candidates for DM velocity substructure, capable of significantly altering the expected modulation \cite{Green,GelminiGondolo,FreeseGondoloNewbergLewis,SavageFreeseGondolo,FornengoScopel,BelliCerulliFornengoScopel,PurcellBullockKaplinghat, ReadLakeAgertzDebattista, BruchReadBaudisLake, ReadMayerBrooksGovernatoLake, FanKatzRandallReece}. Observations of the Sagittarius (Sgr) dwarf galaxy show that tidally stripped stars are currently passing through the galactic plane not far from the Sun. Simulations of this merger suggest that the DM component of the Sgr stream may be passing through the Solar System and could contribute as much as $5\%$ to the local DM density \cite{StiffWidrowFrieman,PurcellZentnerWang}. A DD is a subcomponent of the halo that has a spatial distribution roughly coincident with the visible disk, co-rotating with it, but with a lagging angular velocity \cite{PurcellBullockKaplinghat,ReadLakeAgertzDebattista,BruchReadBaudisLake,ReadMayerBrooksGovernatoLake,FanKatzRandallReece}. Numerical simulations have shown that DDs can form in Milky Way-type galaxies from mergers of satellite galaxies \cite{ReadLakeAgertzDebattista,BruchReadBaudisLake,ReadMayerBrooksGovernatoLake} (although recent measurements suggest this may be unlikely \cite{ReadReview, Ruchti:2015bja}). Alternatively, if a subdominant portion of DM is dissipative in nature it has the potential to collapse and form a DD, a process comparable to the formation of the baryonic disk \cite{FanKatzRandallReece}.

Ref.~\cite{LeeLisantiSafdi} recently performed a Fourier analysis of the expected modulation and considered how the annual and higher harmonics are influenced by the eccentricity of Earth's orbit and the possible existence of DM velocity substructure. Since the eccentricity is small, $e \approx 0.016722$, it is not expected to impact the leading harmonic. DM substructure, however, was shown in Refs.~\cite{LeeLisantiSafdi,SavageFreeseGondolo} to profoundly impact all harmonics, including the unmodulated rate. Furthermore, Ref.~\cite{LeeLisantiSafdi} pointed out that if the DM velocity distribution is smooth and isotropic in the galactic frame, there exist ratios of the amplitudes of the harmonics that are independent of the scattering energy, and thus they concluded that these ratios could be used to probe the level and nature of anisotropy in the DM halo. 

The annual modulation of a DM signal can also be affected by the gravitational focusing (GF) of DM by the Sun \cite{AlenaziGondolo,LeeLisantiPeterSafdi,BozorgniaSchwetz}. The extent to which the annual (first) and biannual (second) harmonics in the SHM are influenced by GF was studied in Refs.~\cite{LeeLisantiPeterSafdi,BozorgniaSchwetz}. They found that GF has a nearly negligible effect on the amplitude of the first harmonic, but can significantly enhance the amplitude of the second harmonic and generate energy-dependent phases in both the first and second harmonics, especially at low energy.

While the Fourier analysis of the rate has been studied for the SHM by taking into account both GF and the eccentricity of Earth's orbit \cite{LeeLisantiPeterSafdi, BozorgniaSchwetz}, the modification of the leading harmonics in the presence of DM substructure has only been studied for an eccentric orbit without GF \cite{LeeLisantiSafdi}. The aim of this paper is to study the effect of GF on a DM halo hosting substructure. We begin by considering the relative importance of GF and the eccentricity of Earth's orbit for the first two harmonics within the SHM, and analyze the extent to which the conclusions of Ref.~\cite{LeeLisantiSafdi} hold when GF is considered. We then study the effect of the Sgr stream and a DD on the annual and biannual harmonics.

In Section \ref{sec:harmonicexpansion} we review the Fourier expansion of the rate and the procedure by which the effect of GF is incorporated. Section \ref{sec:results} presents the amplitudes and phases of the first and second harmonics for the SHM with and without DM substructure. We specifically discuss how GF affects the ability of these harmonics to probe the nature of DM velocity substructure and the extent to which it is present in the galaxy. In Section \ref{sec:numevents} we provide rough estimates of the minimum number of events that would be needed to confirm the existence of an annual modulation for the different halo models we consider, as well as the number of events needed to differentiate between these models. A summary and our conclusions are provided in Section \ref{sec:conclusion}.

\section{DM signal and its modulation}
\label{sec:harmonicexpansion}
Since WIMPs in the galactic halo are non relativistic, $v / c \sim 10^{-3}$, 
the amplitude for DM scattering with a nucleus initially at rest is usually expanded in powers of the small WIMP speed $v$ and momentum transfer $q = \sqrt{2 m_T \ER}$, with $m_T$ the mass of the target nucleus and $\ER$ its recoil energy. The zeroth order term of the scattering amplitude in this expansion is $v$-independent and is usually the only term retained (unless it vanishes). This makes the angular differential WIMP-target nuclide ($T$) cross section $\ud \sigma_T / \ud \cos\theta$ independent of the WIMP speed. However, a more useful quantity entering the analysis of experimental data is the differential cross section in recoil energy $\ud \sigma_T / \ud \ER$.
For elastic scattering, $\ER = v^2 (1- \cos\theta) \mu_T^2 / m_T$, with $\mu_T$ the WIMP-nucleus reduced mass, and one finds $\ud \sigma_T / \ud \ER \propto 1 / v^2$. This proportionality also holds in the leading order term for inelastic scattering.

When the scattering amplitude is independent of $v$, we have $\ud \sigma_T / \ud \ER = \sigma_T(\ER) m_T / (2 \mu_T^2 v^2)$, where $\sigma_T(\ER)$ is a factor with units of a cross section. 
Considering the additional factor of $v$ coming from the DM flux, the scattering rate reads
\begin{equation}
\frac{\ud R_T}{\ud \ER} = C_T \frac{\sigma_T(\ER)}{2 m \mu_T^2} \rho \, \eta(\vmin(\ER), t) \ ,
\label{eq:diffrecoil}
\end{equation}
where $C_T$ is the target mass fraction in the detector, $\rho$ is the local DM mass density, $\vmin(\ER) = \sqrt{m_T \ER / 2 \mu_T^2}$ is the minimum WIMP speed necessary to induce an elastic scattering event with nuclear recoil energy $\ER$, and
\begin{equation}
\eta(\vmin, t) \equiv \int_{v \geqslant \vmin} \frac{f(\bol{v}, t)}{v} \, \ud^3 v 
\label{eq:haloint}
\end{equation}
with $f(\bol{v}, t)$ the local DM velocity distribution in Earth's rest frame. 

To interpret the outcome of direct DM detection experiments within models of particle DM, one typically needs to assume a specific form of the DM velocity distribution. Once this is specified in some reference frame R, \eg the galactic rest frame, the velocity distribution in Earth's rest frame can be obtained by the Galilean transformation $f(\bol{v}, t) = f_\text{R}(\bol{v} + \bol{v}_\text{ES}(t) + \bol{v}_\text{SR})$,
where $f_\text{R}$ is the velocity distribution in R, $\bol{v}_\text{ES}(t)$ is Earth's velocity with respect to the Sun, and $\bol{v}_\text{SR}$ is the Sun's velocity in R. The time dependence of $\bol{v}_\text{ES}(t)$ is due to the annual rotation of Earth about the Sun. For our analysis we take $\bol{v}_\text{ES}(t)$ from Ref.~\cite{LeeLisantiSafdi}.

In the Sun's reference frame, DM particles that are on average at rest with respect to the galaxy are seen to have a preferred direction of motion that opposes $\bol{v}_\text{SR}$. For this reason, the Sun experiences a constant ``wind'' of DM particles. The gravitational potential of the Sun bends the trajectories of these DM particles, acting as a gravitational lens that focuses the DM particles on the leeward side. As a consequence, the DM density and velocity distribution acquire a dependence on Earth's relative location to the Sun (see \eg Fig.~1 of \cite{LeeLisantiPeterSafdi} for a diagrammatic representation).

The effect of GF is taken into account by replacing $f_\text{R}(\bol{v} + \bol{v}_\text{ES}(t) + \bol{v}_\text{SR})$ with $f_\text{R}(\bol{v}_\text{SR} + \bol{v}_\infty[\bol{v}_\text{ES}(t)+\bol{v}])$, where $\bol{v}_\infty[\bol{v}]$ is the velocity a WIMP had very far away from the Sun, where the Sun's gravity is negligible, such that its velocity when it reaches Earth is $\bol{v}$ \cite{AlenaziGondolo,LeeLisantiPeterSafdi}. Ref.~\cite{AlenaziGondolo} has shown that $\bol{v}_\infty[\bol{v}]$ is given by
\begin{equation}
\bol{v}_\infty[\bol{v}] = \frac{v_\infty^2\bol{v} + \frac{1}{2} v_\infty u_\text{esc}^2 \bol{\hat{r}} - v_\infty \bol{v}(\bol{v} \cdot \bol{\hat{r}})}{v_\infty^2 + \frac{1}{2} u_\text{esc}^2 - v_\infty(\bol{v} \cdot \bol{\hat{r}})} \ ,
\label{eq:vinfty}
\end{equation}
where $u_\text{esc} = \sqrt{2 G M_\odot / r} \approx 40$ km/s is the escape velocity of the Solar System at Earth's position, $r$ is the Sun-Earth distance, and $\bol{\hat{r}}$ is the unit vector pointing from the Sun to Earth. Energy conservation ensures that $v_\infty^2 = v^2 - u_\text{esc}^2$.

The velocity integral in \Eq{eq:haloint} can be written as a Fourier series,
\begin{equation}
\eta(\vmin, t) = a_0(\vmin) + \sum_{n = 1}^\infty \big( a_n(\vmin) \cos[n \omega (t - t_0)] + b_n(\vmin) \sin[n \omega (t - t_0)] \big) \ ,
\label{eq:fourier1}
\end{equation}
with $\omega = 2 \pi / \text{year}$ and $t_0$ an arbitrary phase parameter. If Earth's orbit is assumed to be perfectly circular and the DM velocity distribution is isotropic, choosing $t_0$ to be the time at which the speed of Earth with respect to the galaxy is maximum simplifies \Eq{eq:fourier1} by setting all $b_n = 0$. Accounting for astrophysical uncertainties in the velocity of the Local Standard of Rest, the time at which the speed of Earth with respect to the galaxy is maximum occurs somewhere between May $30^\text{th}$ and June $2^\text{nd}$ \cite{BozorgniaGelminiGondolo}.

While the coefficients of \Eq{eq:fourier1} are more easily computed, we find that it is more intuitive and accessible to characterize each harmonic by a single amplitude and a single $\vmin$-dependent phase:
\begin{equation}
\eta(\vmin, t) = A_0(\vmin) + \sum_{n = 1}^\infty A_n(\vmin) \cos[n \omega (t - t_n(\vmin))] \ ,
\label{eq:fourier2}
\end{equation}
with all $A_n \geqslant 0$. Comparing \Eq{eq:fourier2} with \Eq{eq:fourier1} we find
\begin{align}
A_{n}=\sqrt{a_{n}^{2}+b_{n}^{2}} \ ,
&&&
t_n = - \frac{1}{n \omega} \arctan \! \left[ \frac{a_n \sin{(n \omega t_0)} - b_n \cos{(n \omega t_0)}}{a_n \cos{(n \omega t_0)} + b_n \sin{(n \omega t_0)}} \right] .
\label{eq:phasedecomp}
\end{align}
In the next section we compare the amplitudes and phases of the first few Fourier modes of $\eta(\vmin, t)$ for a variety of DM velocity distributions, both with and without GF. These can be computed analytically if one excludes the contribution of GF, otherwise the calculation needs to be done numerically \cite{LeeLisantiPeterSafdi}.

When considering more than one halo component (\eg SHM plus a stream or a DD), we assume that the DM consists of a single type of particle, unless otherwise noted. In this case the amplitude of each mode in the expansion \eqref{eq:fourier1} for the various halo components can be combined linearly, weighted by their density contribution, \ie $a_n^\text{tot}(\vmin) = \sum_i (\rho_i / \rho_\text{tot}) a_{n, i}(\vmin)$ (and analogous equation for $b_n(\vmin)$), where $i$ labels the various DM subcomponents and $\rho_\text{tot} \equiv \sum_i \rho_i$. Notice that the amplitude $A_n$ of the expansion in \Eq{eq:fourier2} is not in general obtained by linearly combining the Fourier amplitudes of different DM subcomponents, since the phases will in general be different. If the DM is assumed to be composed of multiple types of particles with different masses and/or interactions, the combination is not so straightforward and one must take into account the whole factor multiplying $\eta(\vmin, t)$ in \Eq{eq:diffrecoil}. This will be relevant for the case of DM with a dissipative component \cite{FanKatzRandallReece} in Section \ref{subsec:dd}.

\section{Modulation analysis for various halo models}
\label{sec:results}

\subsection{SHM}
\label{subset:shm}

In the SHM the DM velocity distribution in the galactic rest frame is described by an isotropic Maxwellian truncated at the galactic escape speed $\vesc$:
\begin{equation}
f_\text{R}(\bol{v}) = \frac{e^{- \bol{v}^2 / v_0^2}}{\left( \pi v_0^2 \right)^{3/2} N_\text{esc}} \, \Theta(\vesc - |\bol{v}|) \ ,
\label{eq:vdf}
\end{equation}
where $v_0$ is the most probable speed in the galactic rest frame and the normalization is chosen so that $\int \ud^3 v \, f_\text{R}(\bol{v}) = 1$,
\begin{equation}
N_\text{esc} = \mathrm{Erf}(\vesc / v_0) - \frac{2 \vesc}{\sqrt{\pi} v_0} e^{- \vesc^2 / v_0^2} \ .
\label{eq:normfv}
\end{equation}
We assume $v_0$ to be equal to the speed of the Local Standard of Rest, $v_0 = 220$ km/s \cite{Kerr}, we take the Sun's velocity with respect to the galaxy from Ref.~\cite{LeeLisantiSafdi}, $\bol{v}_\text{SR}  = (11, 232, 7)$ km/s, and $\vesc = 533$ km/s \cite{RAVE}.

\begin{figure*}
\includegraphics[width=\textwidth]{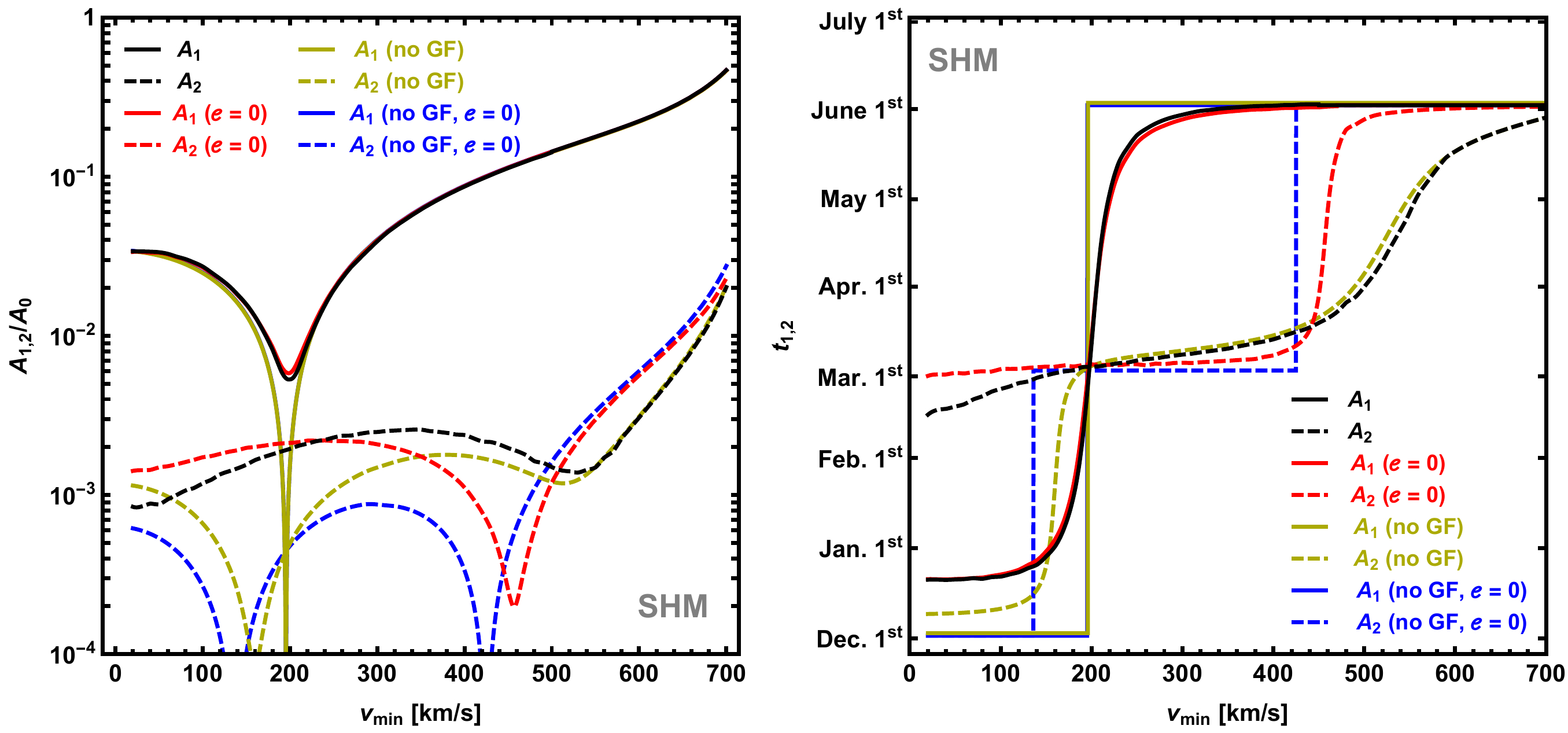}
\caption{Amplitudes (left) and phases (right) for the first (solid) and second (dashed) harmonics, for the SHM including the effect of GF and the eccentricity of Earth's orbit (black), including GF but neglecting the eccentricity (red), including the eccentricity but neglecting GF (yellow), and neglecting both GF and the eccentricity (blue). The eccentricity has a negligible impact on the first harmonic, therefore the solid black line extensively overlaps the red, and the solid dark yellow line completely overlaps the blue, for both the amplitude and the phase.}
\label{fig:SHM}
\end{figure*}

Fig.~\ref{fig:SHM} shows the amplitudes and phases of the first (solid lines) and second (dashed lines) harmonic including both GF and the eccentricity of Earth's orbit in the calculations (black), including the eccentricity but neglecting GF (yellow), including GF but neglecting the eccentricity (red), and neglecting both the eccentricity and GF (blue). Notice that, due to the negligible impact of the eccentricity on the first harmonic, the solid black line extensively overlaps the red, and the solid dark yellow line completely overlaps the blue, for both the amplitude and the phase.
In agreement with Refs.~\cite{LeeLisantiPeterSafdi,BozorgniaSchwetz}, Fig.~\ref{fig:SHM} shows that the inclusion of GF  causes an approximate 20 day shift in the phase of the annual harmonic for $\vmin \lesssim 100$ km/s and eliminates the sudden phase flips, \ie the occurrence of a jump in the phase of a given harmonic by half the period. The phase flip can also be identified by the vanishing of the amplitude of a given harmonic. When GF is included, the amplitudes no longer vanish and the phases develop a softer $\vmin$ dependence. Actually, the presence of any anisotropy eliminates the phase flip and leads to continuous transitions in the phase. For the remainder of this paper we will loosely use the term phase flip to refer to both the previously defined jump in the phase, and the rapid, but continuous, phase transitions that may appear when anisotropy is present. 

\begin{figure}[t!]
\centering
\includegraphics[width=0.5\textwidth]{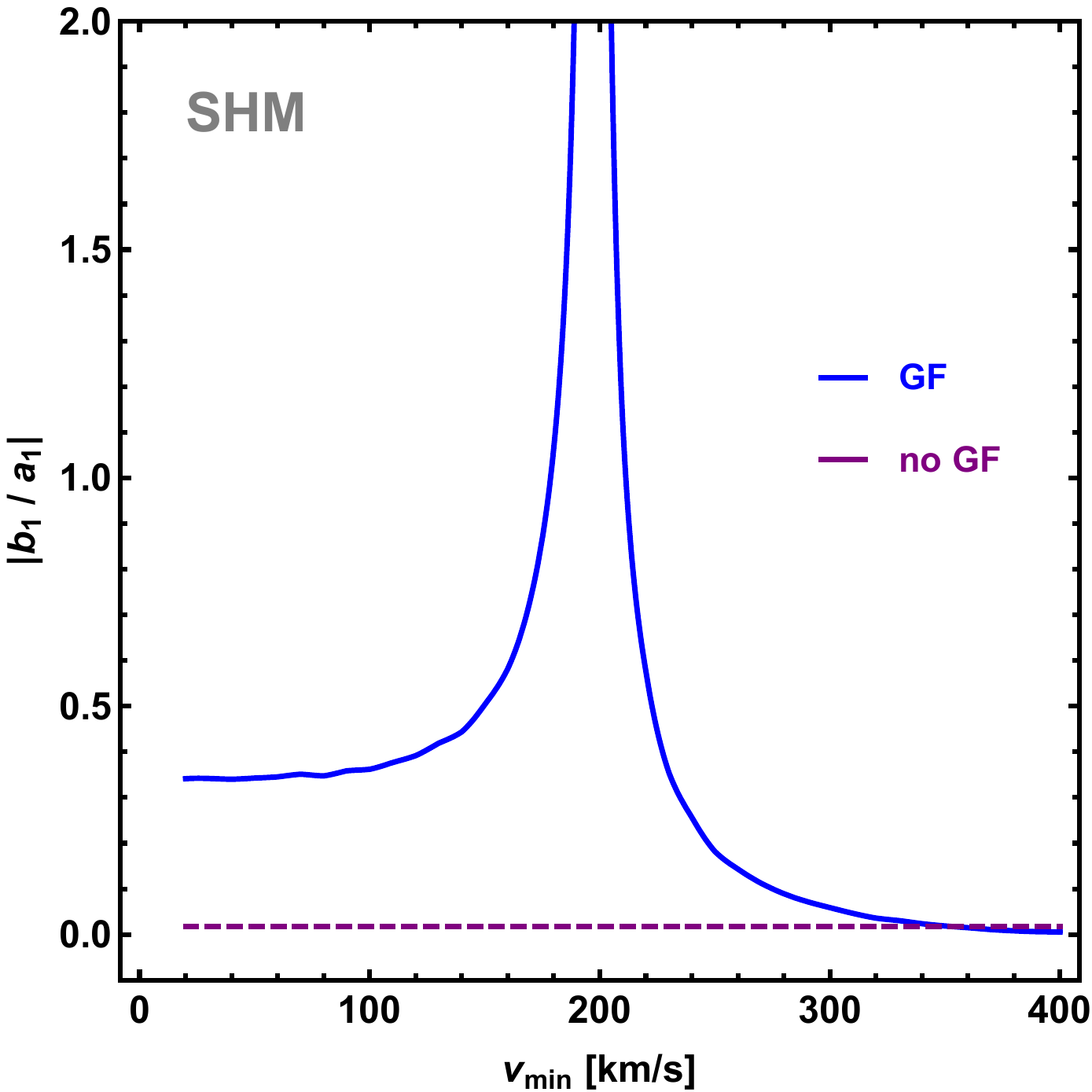}
\caption{Ratio $|b_1 / a_1|$ (see \Eq{eq:fourier1}) for the SHM, with GF (solid blue line) and without GF (dashed purple line). Without GF this ratio is approximately $1/59$.}
\label{fig:SHMb1a1}
\end{figure}

The conclusion of Ref.~\cite{LeeLisantiSafdi} that the ratios of the amplitudes of various harmonics can be used to probe the anisotropy of the DM halo was based on the assumption that the DM velocity distribution is isotropic in the galactic frame. The existence of GF is not consistent with this assumption. Thus the ratios in Ref.~\cite{LeeLisantiSafdi} only hold for large $\vmin$, where the effect of GF is not significant because DM particles spend little time in Sun's gravitational potential. Fig.~\ref{fig:SHMb1a1} shows the ratio of coefficients $b_1$ and $a_1$, defined in \Eq{eq:fourier1}, with GF (solid blue line) and without GF (dashed purple line). One can see that without GF this ratio would be independent of the scattering energy and remains at a constant value of $\simeq1/59$, as found in Ref.~\cite{LeeLisantiSafdi}. GF significantly alters this result for $\vmin \lesssim 300$ km/s.

\subsection{Sagittarius Stream}
\label{subsec:sgr}

Since the parameters governing the velocity and dispersion of the Sgr stream are not well known, we will follow the assumptions of \cite{FreeseGondoloNewbergLewis}. In the galactic coordinate system, we take the mean velocity of the Sgr stream in the galactic frame to be $\bol{v}_\text{Sgr} = (-65, 135, -249)$ km/s, and model this component with an isotropic Maxwellian with dispersion $v_0 = 30 \sqrt{2/3} \text{ km/s} \approx 25$ km/s. We model the smooth virialized component of the halo with the SHM.

Ref.~\cite{PurcellZentnerWang} recently studied the effect of the Sgr stream with a self-consistent N-body simulation and found that the addition of the stream can noticeably alter $\eta(\vmin, t)$. Specifically, they found four major changes. First, incorporating the Sgr stream in a realistic halo model with a baryonic disk produced a $10$--$20\%$ increase in the direct search event rate for values of $\vmin$ larger than the typical relative stream speed.
Additionally, Ref.~\cite{PurcellZentnerWang} found a $20$--$30\%$ decrease in the fractional modulation amplitude defined as
\begin{equation}
\Delta \equiv \frac{\eta_\text{max} - \eta_\text{min}}{\eta_\text{max} + \eta_\text{min}} \ ,
\label{eq:maxamp}
\end{equation}
where $\eta_\text{max}(\vmin)$ and $\eta_\text{min}(\vmin)$ are the maximum and minimum of $\eta(\vmin, t)$ in time. If the modulation is perfectly sinusoidal with a period of one year, $\Delta$ coincides with the amplitude of the sinusoid normalized by the unmodulated component of $\eta(\vmin, t)$, \ie $A_1 / A_0$. Furthermore, the phase flip of the dominant harmonic was found to occur about $10$--$15$ km/s lower in $\vmin$, and deviations of up to about $20$ days were found in the phase of the modulation at values of $\vmin$ near the typical speed of a WIMP belonging to the stream as seen in the Sun's reference frame. We begin by investigating if and how these conclusions change when GF is taken into account.

\begin{figure*}
\centering
\includegraphics[width=\textwidth,trim=0mm 10mm 0mm 0mm, clip]{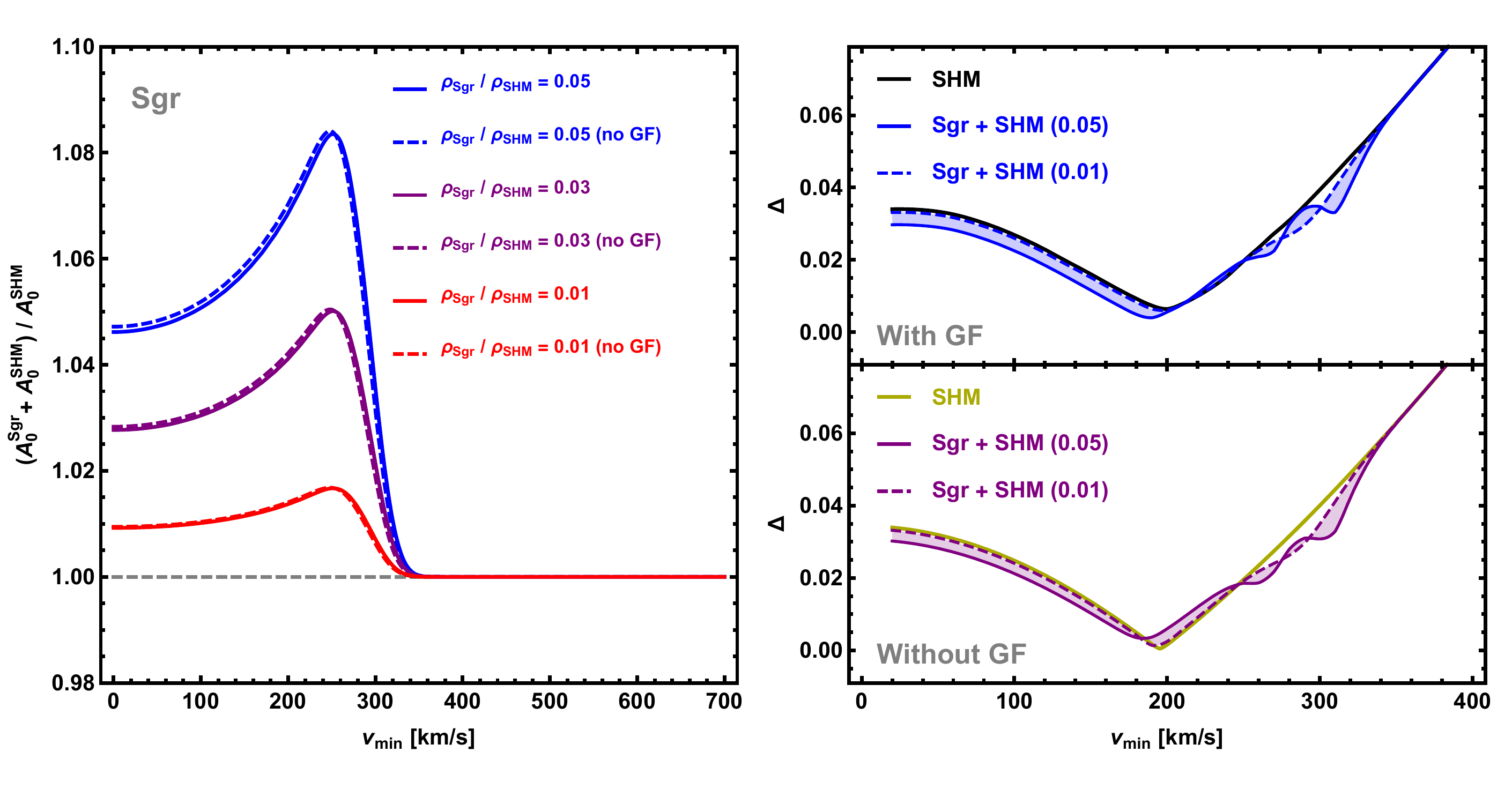}
\caption{Left: Unmodulated component of $\eta(\vmin, t)$, $A_0$, for Sgr+SHM, normalized by the unmodulated component for the SHM alone, for $\rho_\text{Sgr} / \rho_\text{SHM} = 0.01$ (red), $ \rho_\text{Sgr} / \rho_\text{SHM} = $ $0.03$ (purple), and $ \rho_\text{Sgr} / \rho_\text{SHM} = $ $0.05$ (blue), with GF (solid) and without GF (dashed). Right: Fractional modulation amplitude, as defined in \Eq{eq:maxamp}, for Sgr+SHM with GF (top panel) and without GF (bottom panel), and for stream densities ranging from $\rho_\text{Sgr} / \rho_\text{SHM} = 0.01$ (dashed) to $ \rho_\text{Sgr} / \rho_\text{SHM} = $ $0.05$ (solid). Results are compared with the SHM alone, with GF (black) and without GF (yellow).}
\label{fig:sgrcomp}
\end{figure*}

The left panel of Fig.~\ref{fig:sgrcomp} contains the relative increase in the unmodulated component of $\eta(\vmin, t)$ when the Sgr stream is added to the SHM (Sgr+SHM), for stream densities ranging from 1\% to 5\% of the DM halo density. While Fig.~\ref{fig:sgrcomp} does show an increase in the unmodulated component, GF appears to have no additional effect. The reason for this is explained in further detail in the following paragraphs.   

The right panels of Fig.~\ref{fig:sgrcomp} show the fractional amplitude as defined in \Eq{eq:maxamp} for the Sgr+SHM, for $\rho_\text{Sgr} / \rho_\text{SHM} = 0.01$ (dashed) and $0.05$ (solid), with (top) and without (bottom) GF. Here and in the following, the shaded regions between the dashed and solid lines highlight where lines corresponding to intermediate densities lie. As with the unmodulated component of the rate, the fractional modulation amplitude for Sgr+SHM seems to not be affected by GF. Small deviations from the SHM do occur at $\vmin \lesssim 200$ km/s, and at values of $\vmin$ near the speed of the stream in Earth's frame, regardless of GF. However, Fig.~4 of Ref.~\cite{PurcellZentnerWang} shows that the uncertainty in the velocity distribution of the virialized component of the dark halo, specifically deviations from the assumed Maxwellian distribution, have a larger influence on $\Delta(\vmin)$ than the addition of the Sgr stream to the SHM. 

\begin{figure*}
\centering
\includegraphics[width=\textwidth,trim=0mm 15mm 0mm 0mm, clip]{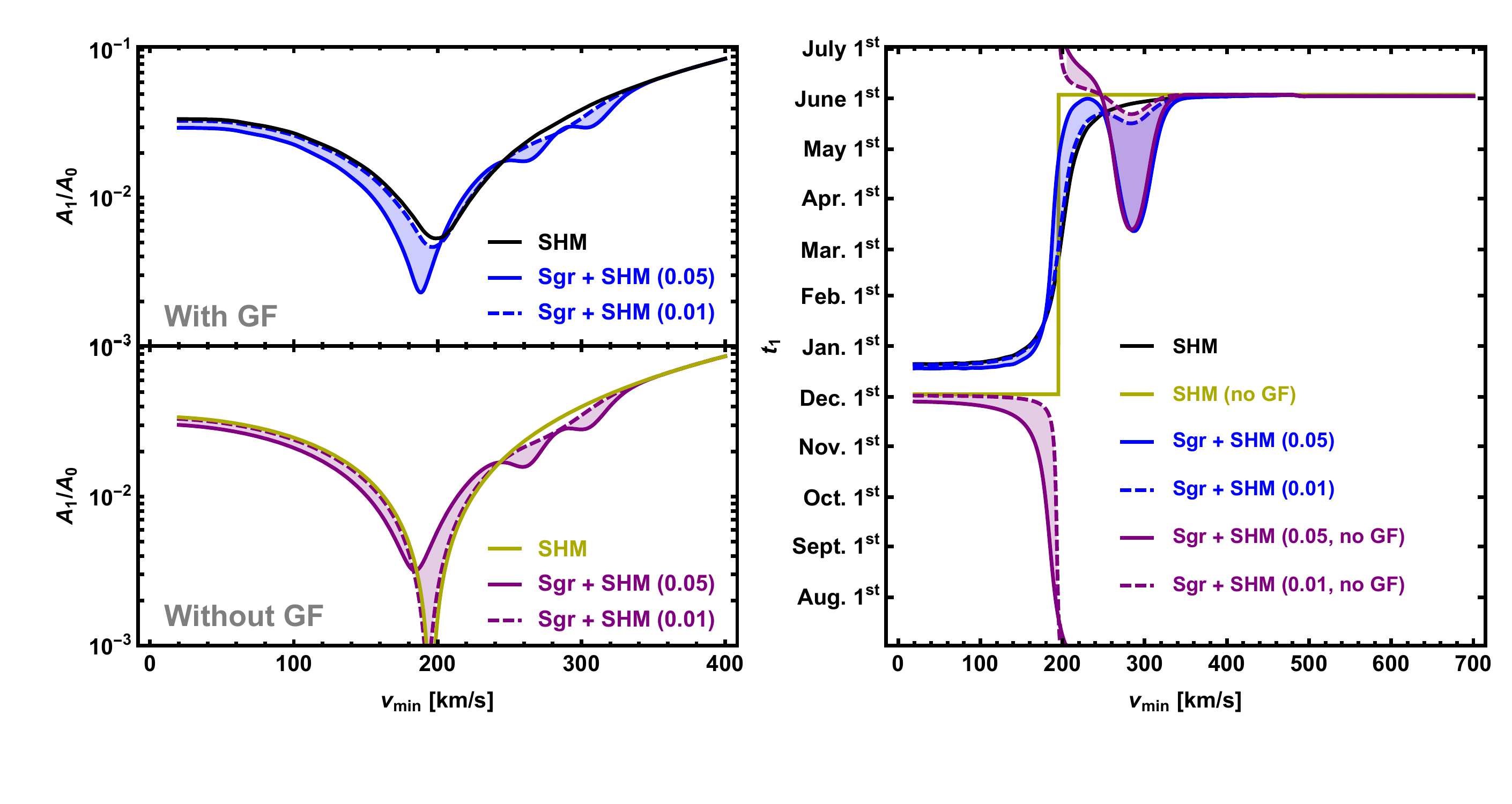}
\caption{Amplitudes (left) and phases (right) of the first harmonic for the Sgr+SHM. Dashed lines are for a low density stream, $ \rho_\text{Sgr} / \rho_\text{SHM} = 0.01$, while solid lines are for a high density stream, $ \rho_\text{Sgr} / \rho_\text{SHM} = 0.05$. Blue lines and regions include GF, while purple lines and regions neglect GF. Results are compared with the SHM alone, with GF (black) and without GF (yellow).}
\label{fig:sgr1}
\end{figure*}

\begin{figure*}
\centering
\includegraphics[width=\textwidth,trim=0mm 15mm 0mm 0mm, clip]{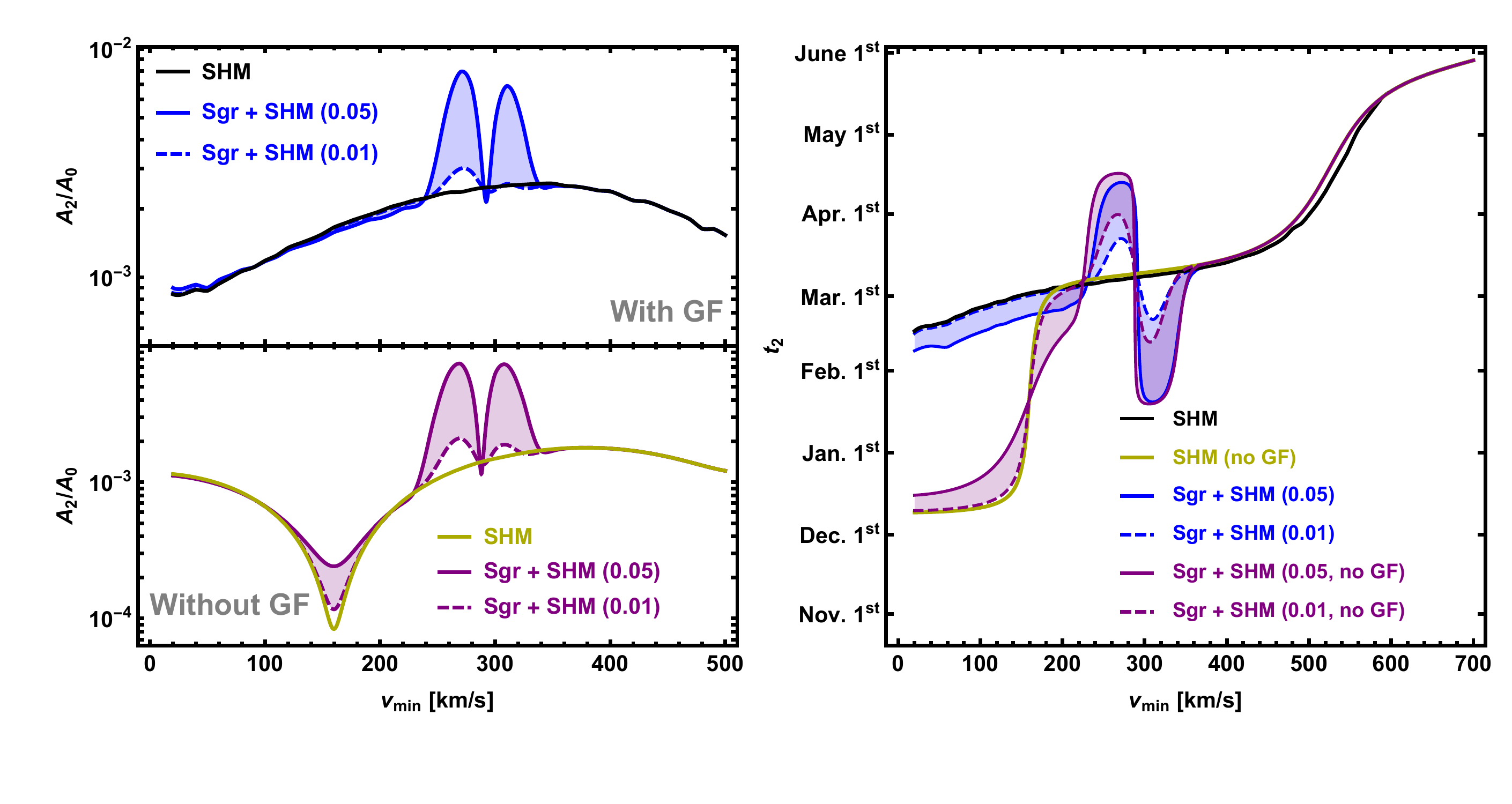}
\caption{Same as Fig.~\ref{fig:sgr1}, but for the second harmonic.}
\label{fig:sgr2}
\end{figure*}

The effect of GF is known to increase as the relative WIMP velocity in the solar frame decreases. This is because slower WIMPs spend more time in the Sun's gravitational potential. The parameters of the Sgr stream chosen in this paper imply the majority of the WIMPs coming from the Sgr stream move at roughly $v_\text{Sgr}^\text{S} \equiv 300$ km/s in the Sun's frame, with very few traveling below 250 km/s. Thus we do not find it surprising that GF has little impact on the unmodulated component of $\eta(\vmin,t)$ and the fractional modulation amplitude when the Sgr stream is added to the SHM. Notice that the SHM component of the Sgr+SHM has a lower average WIMP speed and a much larger dispersion with respect to the Sgr component, implying WIMPs from the background component of the dark halo will be more affected by GF than those in the stream. Figs.~\ref{fig:sgr1} and \ref{fig:sgr2} show how this influences the amplitudes and phases of the first and second harmonics.

Fig.~\ref{fig:sgr1} shows the amplitude $A_1$ (left) and the phase $t_1$ (right) of the first harmonic. Without GF, $A_1$ would experience almost no deviation from the SHM, except around the $\vmin$ values where the SHM has a dip due to the phase flip, and in a small region near $v_\text{Sgr}^\text{S}$ where it decreases by at most a factor of $2$. Including GF does not noticeably affect these results. When GF is accounted for, the flip of $t_1$ in the Sgr+SHM is delayed relative to the SHM alone, but it occurs more rapidly, causing a deviation in the phase by up to two months.
As found in Ref.~\cite{PurcellZentnerWang}, when the Sgr stream is added to the smooth component of the halo there appears to be a significant deviation of up to two months in the phase of the annual modulation near $\vmin \simeq v_\text{Sgr}^\text{S}$. This effect occurs regardless of GF, and the inclusion of GF even appears to enhance this deviation for low density streams, resulting in an approximately 20 day phase shift for $\rho_\text{Sgr} / \rho_\text{SHM} = 0.01$.

Fig.~\ref{fig:sgr2} shows the amplitude $A_2$ (left) and the phase $t_2$ (right) of the second harmonic. The results for the phase of the second harmonic are very similar to those of the first. Without GF, $A_2$ would deviate significantly from the SHM only at $\vmin \simeq v_\text{Sgr}^\text{S}$ and around the point where the SHM amplitude has a dip due to a phase flip. When GF is included the dip vanishes but the enhancement of the amplitude at  $\vmin \simeq v_\text{Sgr}^\text{S}$ remains. This enhancement is roughly a factor of $4$ for $\rho_\text{Sgr} / \rho_\text{SHM} = 0.05$.  Without GF, the phase would exhibit a flip at smaller $\vmin$ values and would experience strong deviations from the SHM of up to $\sim 50$ days for $250 \text{ km/s} < \vmin < 350$ km/s. The inclusion of GF again washes out the low $\vmin$ deviations, but leaves those above $\vmin \approx 200$ km/s intact. Thus one would expect a $1\%$ density stream to have a second harmonic phase similar to the SHM's, except in the region $\vmin \approx v_\text{Sgr}^\text{S}$, where the stream could cause deviations of as much as 45 days.

To summarize, including GF in Sgr+SHM calculations does not affect the unmodulated rate or the fractional amplitude. Furthermore, it appears to wash out characteristics that may allude to the potential existence of DM substructure in the phases of the first and second harmonics for values of $\vmin \lesssim 200$ km/s. The features arising from the substructure near $\vmin \approx v_\text{Sgr}^\text{S}$, however, are left intact. Since the effect of GF is negligible at $\vmin \gtrsim 300$ km/s, should the Sgr stream contribute non-negligibly to the local DM density, anisotropies arising from the Sgr stream could be probed using the ratios of harmonics as proposed in Ref.~\cite{LeeLisantiSafdi}.

\subsection{Dark Disk}
\label{subsec:dd}

We consider here two distinct types of DDs. The first could form from accretions of massive satellites onto the galactic disk \cite{PurcellBullockKaplinghat,ReadLakeAgertzDebattista,BruchReadBaudisLake,ReadMayerBrooksGovernatoLake}. In this scenario, the DM in the halo of the satellite galaxies and the DM comprising our own galaxy's halo is expected to be of the same type, \ie non-dissipative in nature. While the DM in the halo must be non-dissipative in order to maintain its known spatial distribution, Ref.~\cite{FanKatzRandallReece} has shown that a subdominant portion of at most $5\%$ could be dissipative. If such a component exists, it would form a DD in much the same way the baryonic matter dissipates energy and forms the visible disk. We refer to this second scenario as dissipative dark disk (DDD). Densities of DDs are expected to range from $\rho_\text{disk} / \rho_\text{halo} = 0.2$ to $1$, with this ratio being strictly less than $3$ and likely less than $2$ \cite{BruchReadBaudisLake}.

We model both the non-dissipative DD and the DDD using a truncated Maxwellian. Consistent with the values found in numerical simulations, we model the DD with a rotation velocity $50$ km/s slower than the Local Standard of Rest ($v_\text{lag}=50$ km/s) and with velocity dispersion $v_0 = 70$ km/s \cite{ReadLakeAgertzDebattista}. While we present our results for a non-dissipative DD together with a SHM halo (DD+SHM), results for the DDD are obtained without a background halo component, as a specific particle model for all (dissipative and non-dissipative) DM components would be required before the velocity integrals of SHM and DDD could be combined.

\begin{figure*}
\centering
\includegraphics[width=\textwidth,trim=0mm 10mm 0mm 0mm, clip]{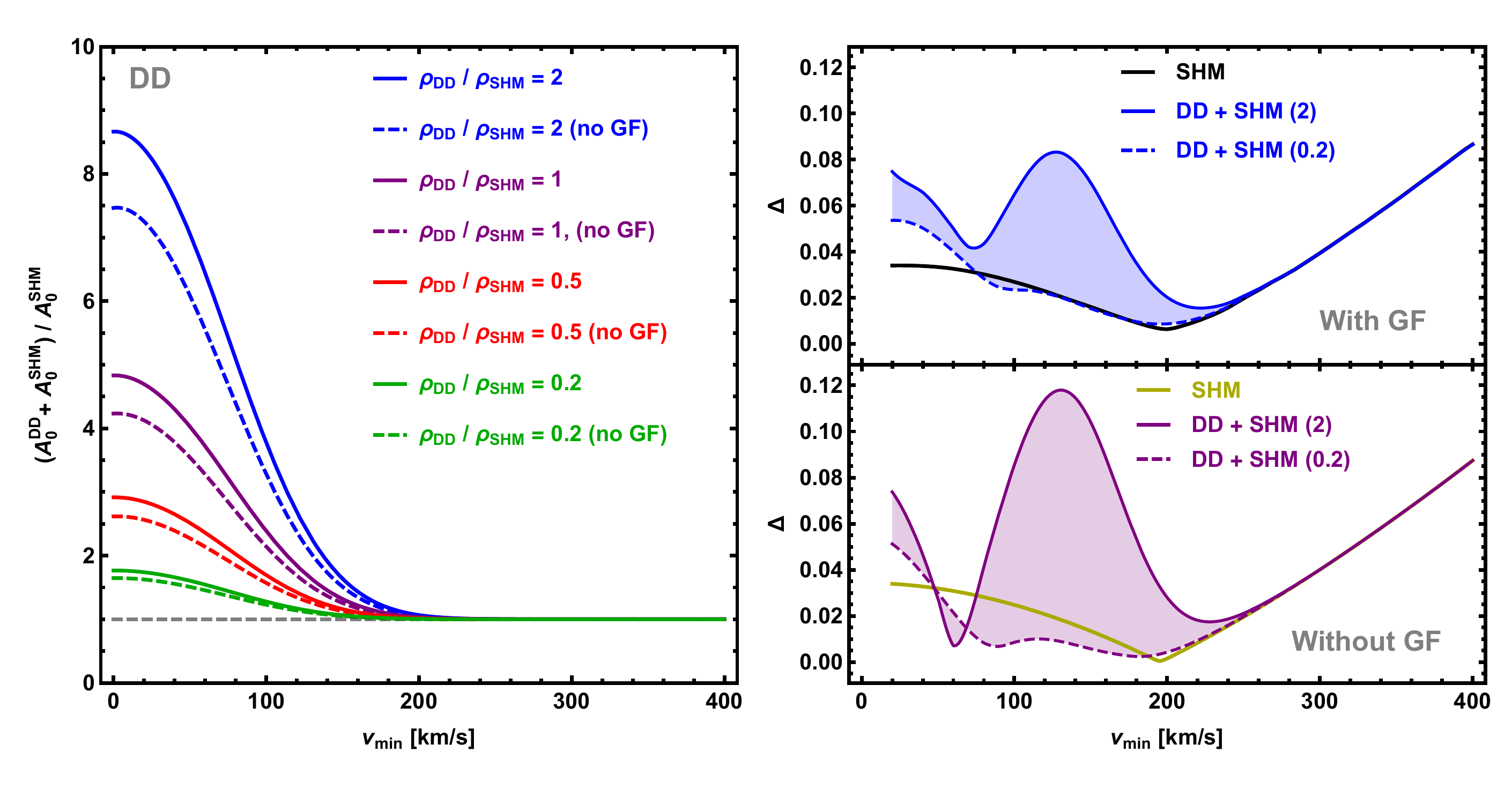}
\caption{Left: Unmodulated component of $\eta(\vmin, t)$, $A_0$, for the DD+SHM normalized by the unmodulated component for the SHM alone, for $\rho_\text{DD} / \rho_\text{SHM} = 0.2$ (green), $ \rho_\text{DD} / \rho_\text{SHM} = 0.5$ (red), $ \rho_\text{DD} / \rho_\text{SHM} = 1$ (purple), and $ \rho_\text{DD} / \rho_\text{SHM} = 2$ (blue), with GF (solid) and without GF (dashed). Right: Fractional modulation amplitude, as defined in \Eq{eq:maxamp}, for the DD+SHM with (top) and without (bottom) GF, and densities ranging from $\rho_\text{DD} / \rho_\text{SHM} = 0.2$ (dashed) to $ \rho_\text{DD} / \rho_\text{SHM} = 2$ (solid). Results are compared with the SHM alone, with GF (black) and without GF (yellow).}
\label{fig:ddcomp}
\end{figure*}

The left panel of Fig.~\ref{fig:ddcomp} shows the enhancement of the unmodulated component of $\eta(\vmin, t)$, $A_0$, for a DD combined with the SHM (DD+SHM) relative to that of the SHM alone, for $\rho_\text{DD} / \rho_\text{SHM} = 0.2$ (green), $0.5$ (red), $1$ (purple), and $2$ (blue), with (solid) and without (dashed) GF. When GF is neglected, adding a DD to the halo can increase the unmodulated rate by as little as $150\%$ for $\rho_\text{DD} / \rho_\text{SHM} = 0.2$, or as much as $775\%$ for $\rho_\text{DD} / \rho_\text{SHM} = 2$, but only for $\vmin \leqslant 200$ km/s. GF does not appreciably change this result for the low density DD, but can increase the unmodulated component of the high density DD by an additional $60\%$. 

The right panel of Fig.~\ref{fig:ddcomp} shows the fractional modulation amplitude $\Delta(\vmin)$ for the DD+SHM, defined in \Eq{eq:maxamp}. For values of $\vmin$ below 300 km/s, where one expects GF and a DD with small $v_\text{lag}$ to be most influential, the SHM predicts a fractional modulation amplitude of at most $\sim 4\%$. Without GF, the addition of the DD to the SHM would increase $\Delta$ to as much as $12\%$. The presence of GF reduces the influence of the DD, resulting in at most an $8\%$ fractional modulation amplitude. For $\vmin > 250$ km/s, the influence of the DD and GF vanish and the DD+SHM results are identical to those of the SHM.

\begin{figure*}
\centering
\includegraphics[width=\textwidth,trim=0mm 15mm 0mm 0mm, clip]{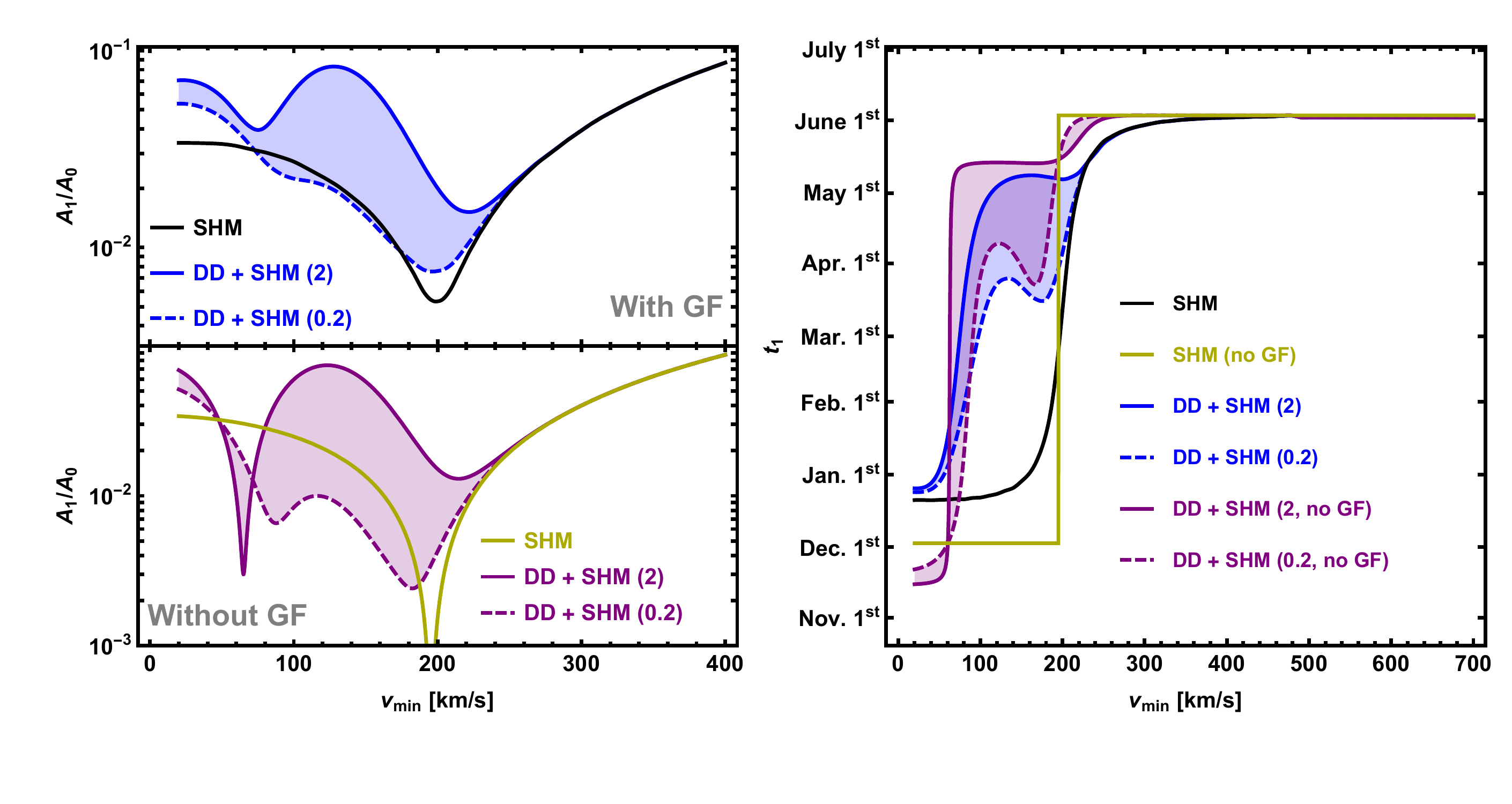}
\caption{Amplitudes (left) and phases (right) of the first harmonic for the DD+SHM. Dashed lines are for a low density DD, $ \rho_\text{DD} / \rho_\text{SHM} = 0.2$, while solid lines are for a high density DD, $ \rho_\text{DD} / \rho_\text{SHM} = 2$. Blue lines and regions include GF, while purple lines and regions neglect GF. Results are compared with the SHM alone, with GF (black) and without GF (yellow).}
\label{fig:dd1}
\end{figure*}

The left panel of Fig.~\ref{fig:dd1} shows the amplitude of the first harmonic. Without GF (purple), the addition of the DD would either increase or decrease the relative amplitude of the first harmonic, depending on the values of $\rho_\text{DD} / \rho_\text{SHM}$ and $\vmin$. Due to GF the addition of the DD primarily enhances the amplitude of the first harmonic. This enhancement can be as large as a factor of 5 for $\rho_\text{DD}/\rho_\text{SHM} = 2$.

The right panel of Fig.~\ref{fig:dd1} shows how the DD impacts the expected phase of the first harmonic, with (blue) and without (purple) GF. Regardless of whether or not GF is included, the phase of the first harmonic of the DD+SHM looks identical to that of the SHM for $\vmin > 250$ km/s. Without GF, the phase of the first harmonic would deviate from the SHM by nearly $6$ months for $\vmin$ between $70$ km/s and $200$ km/s. When GF is accounted for, this phase difference between the DD+SHM and SHM is at most $4$ months, and the range of $\vmin$ at which this deviation occurs is reduced to $100$ km/s $\leqslant \vmin \leqslant 180$ km/s.

\begin{figure*}
\centering
\includegraphics[width=\textwidth,trim=0mm 15mm 0mm 0mm, clip]{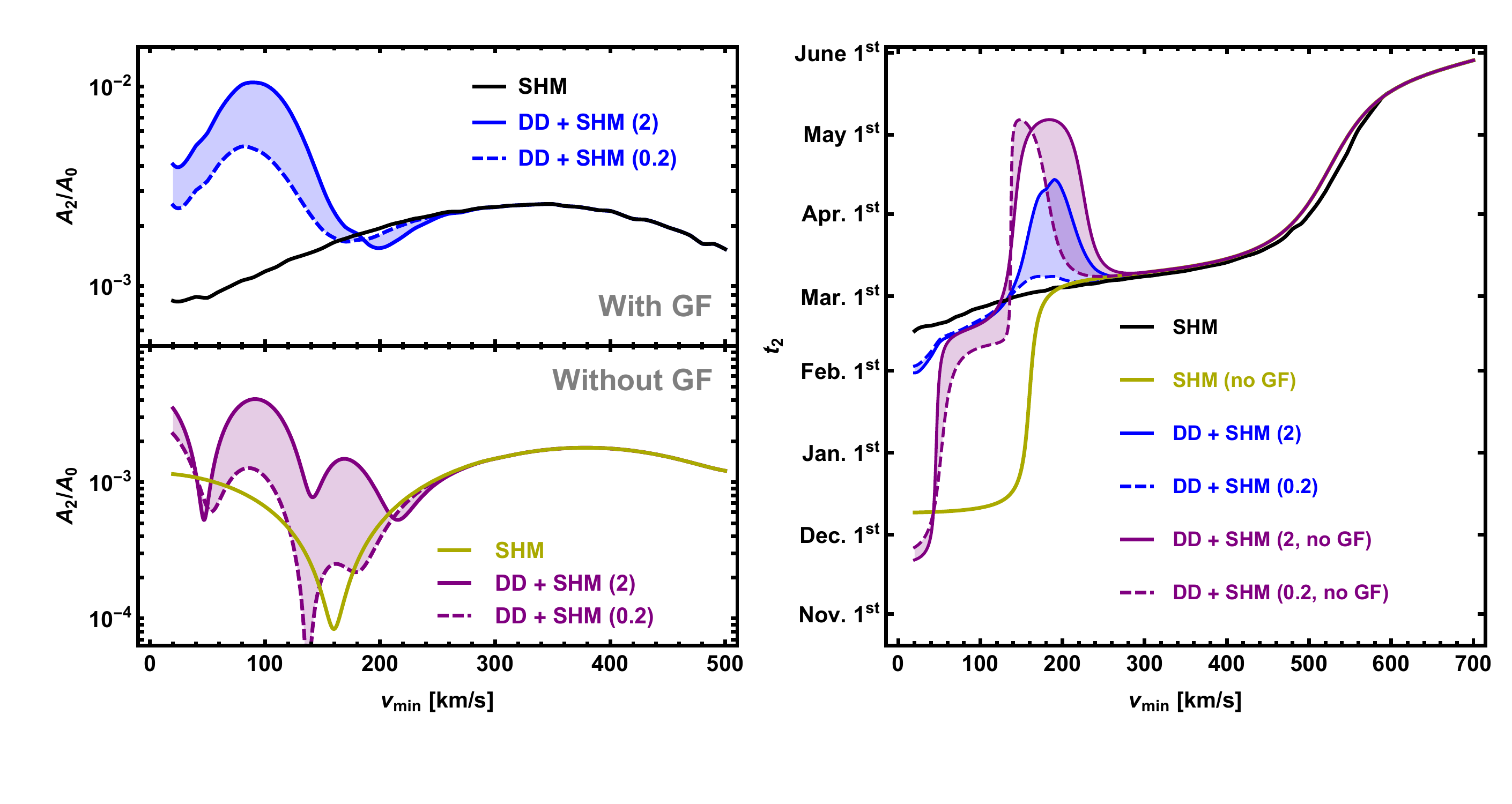}
\caption{Same as Fig.~\ref{fig:dd1} but for the second harmonic.}
\label{fig:dd2}
\end{figure*}

Fig.~\ref{fig:dd2} shows the effect of a DD on the second harmonic. As for the first harmonic, the DD affects neither the relative amplitude nor the phase for $\vmin > 250$ km/s. Below this value, the relative amplitude of the second harmonic of the DD+SHM is primarily enhanced when GF is neglected, except in very narrow regions around $\vmin \simeq 50$ and $\vmin \simeq 140$ km/s, depending on $\rho_\text{DD}$. When GF is included, the amplitude for the DD+SHM is enhanced by up to a factor of 10 for $\rho_\text{DD}/\rho_\text{SHM} = 2$ and a factor of 3 for  $\rho_\text{DD}/\rho_\text{SHM} = 0.2$, but only at values of $\vmin$ below $180$ km/s. A slight reduction in the amplitude occurs for all DD densities plotted between $\vmin$ values of $180$ km/s and $250$ km/s. The phase of the second harmonic without GF would consistently differ from the SHM by up to $75$ days for values of $\vmin$ below $250$ km/s. GF slightly reduces the difference in $t_2$ between DD+SHM and SHM alone for $\vmin < 250$ km/s. The maximum phase difference between DD+SHM and the SHM is roughly $50$ days, but may be as little as $15$ days for $\rho_\text{DD} / \rho_\text{SHM} = 0.2$.

In summary, we find that the existence of a DD with a lag speed of $50$ km/s and dispersion of $70$ km/s can significantly alter the unmodulated rate, the fractional modulation amplitude, and the phases of the dominant harmonics, but only for $\vmin \lesssim 250$ km/s. A larger (smaller) $v_\text{lag}$ would increase (decrease) the $\vmin$ values at which the features associated with the DD appear. GF is shown to further enhance the unmodulated rate and fractional amplitude, but it diminishes the influence of the DD on the phases of the dominant harmonics. However, one should keep in mind that the relative importance of GF seen in this paper is dependent upon the chosen rotation velocity of the DD. 

We notice that the method of Ref.~\cite{LeeLisantiSafdi} for probing the nature of anisotropies in the dark halo with ratios of harmonics may not be useful for a DD, as the region in which deviations occur directly overlaps with the region made anisotropic by GF (at least in the example provided here). The effect of GF would be noticeably reduced should the DD rotate at a velocity significantly different from the rotation velocity of the Sun. This is because, in the Sun's reference frame, WIMPs from the DD will be moving faster and spend less time in the Sun's gravitational potential. 

\begin{figure*}
\centering
\includegraphics[width=\textwidth,trim=0mm 10mm 0mm 0mm, clip]{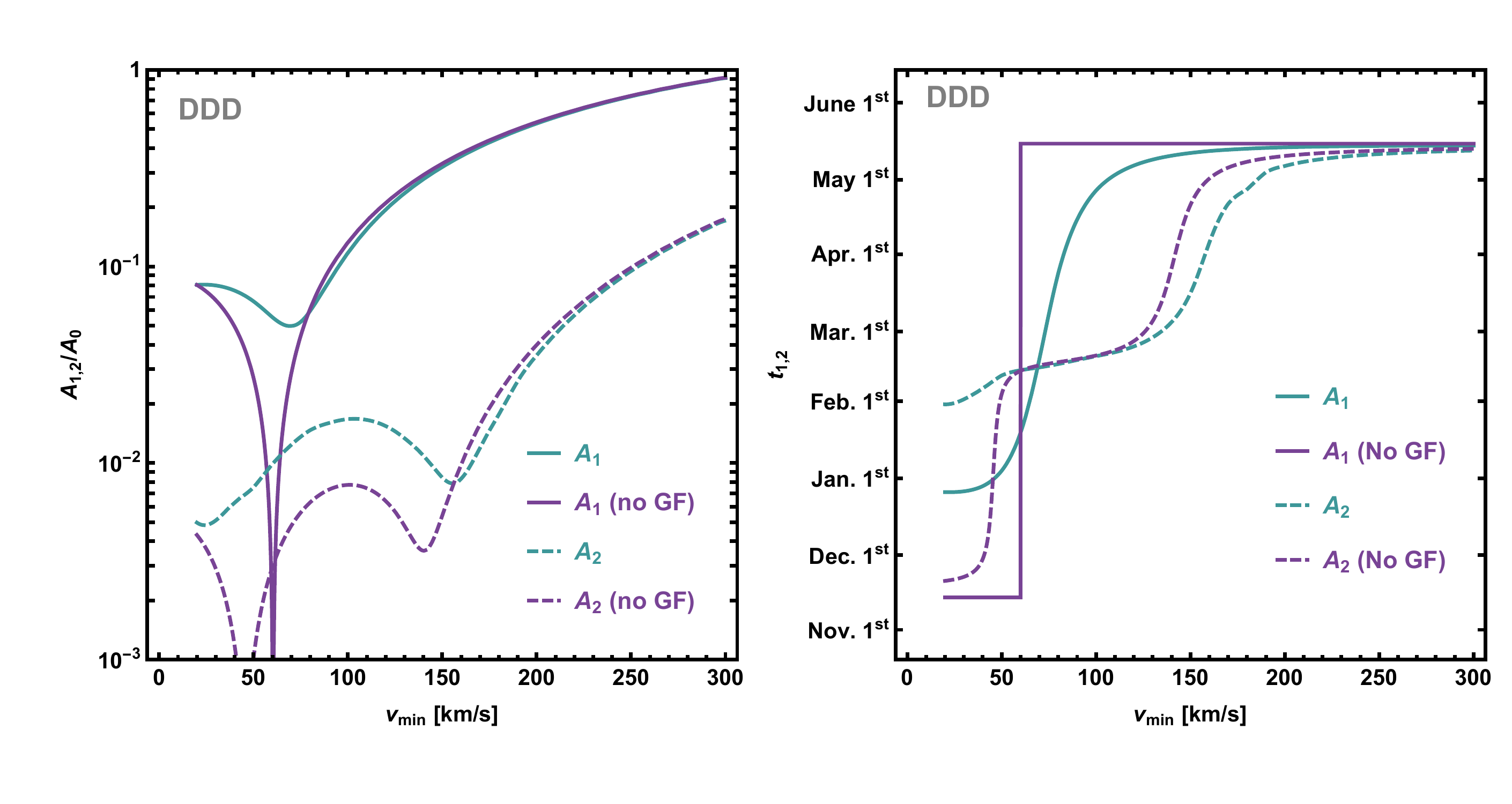}
\caption{Amplitudes (left panel) and phases (right panel) of the first (solid) and second (dashed) harmonics for the DDD, with GF (light blue) and without GF (purple).}
\label{fig:dddm}
\end{figure*}

We also show in Fig.~\ref{fig:dddm} the amplitudes (left panel) and phases (right panel) of the first (solid lines) and second (dashed lines) harmonics for the DDD alone, with (light blue) and without (purple) GF. The amplitudes for the DDD look very similar to those of the SHM, but they appear at much lower $\vmin$ values for the DDD lag speed we assume ($50$ km/s). This should be expected, as the DDD contains WIMPs coming from approximately the same direction as the dark halo but with a lower relative speed and a smaller dispersion. The most notable differences occur in the phases. For all values of $\vmin$, the phases of the first and second harmonic for the DDD are shifted approximately half a month earlier when compared to the phases of the SHM (see \Fig{fig:SHM} for comparison).

\section{Estimate of required number of events}
\label{sec:numevents}

We begin by providing a rough estimate of the number of events that would be necessary to observe the annual modulation in the Sgr+SHM, the DD+SHM, and the SHM, using a very simple two-bin analysis. Let us split an annual cycle into two six-month periods, one of which is centered about the time of maximum of the rate and the other is centered $6$ months later. For a fixed energy range, we denote the number of events in the two time bins $N_+$ and $N_-$. To estimate the number of events needed to establish the existence of an annual modulation with a significance level corresponding to $\alpha$ standard deviations, we require that 
\beq
\Delta N \equiv N_{+}-N_{-} \geqslant \alpha \sqrt{N_{\text{tot}}} \ ,
\label{eq:deltan}
\eeq
with $N_\text{tot} \equiv N_+ + N_-$ the total number of observed events. We assume that the uncertainty of $N_+$ and $N_-$ is $\sqrt{N_\text{tot}/2}$. Assuming the phase $\bar{t}$ is constant in the energy range considered, we can approximate the integrated rate as $R(t) \simeq R_0+R_1\cos{(2\pi (t-\bar{t}) / \text{year})}$, where $R_0$ is the unmodulated component of the rate and $R_1$ is the modulation amplitude. For a fixed exposure $MT$, $N_{\pm} \simeq MT(R_0/2\pm R_1/\pi)$, where the factor of $1/\pi$ arises from integrating the cosine term over the temporal region defining each bin. 

Solving \Eq{eq:deltan} for $N_\text{tot}$ in terms of $R_0$ and $R_1$ then yields
\beq
N_\text{tot} \geqslant \frac{\alpha^2 \pi^2}{4} \left( \frac{R_0}{R_1} \right)^2 \ .
\label{eq:ntot}
\eeq
$R_0$ and $R_1$ can be replaced by the integral of $A_0$ and $A_1$ over the energy range considered, since all additional constants relating $R_0$ to $A_0$ and $R_1$ to $A_1$ cancel in the ratio. $A_0$ and $A_1$ are functions of $\vmin$ and we use the relation between $\vmin$ and $\ER$ for elastic scattering given below \Eq{eq:diffrecoil}.

Since we would ultimately like to know how distinguishable the Sgr+SHM and DD+SHM are from the SHM, we choose to evaluate \Eq{eq:ntot} in an energy range where the amplitude and phase of the Sgr+SHM and DD+SHM deviate most strongly from the SHM. For the Sgr+SHM, this region corresponds to $\vmin$ values between $280$ km/s and $300$ km/s (see \Fig{fig:sgr1}). This region is roughly consistent with a $1$ keV bin centered at about $6.6$ keV, for a $25$ GeV DM particle scattering off xenon. Evaluating \Eq{eq:ntot} in this region we find the modulation amplitude in the Sgr+SHM, for $\rho_\text{Sgr} / \rho_\text{SHM} = 0.05$, requires roughly $2900\alpha^2$ events to be detected with significance $\alpha$ sigma, while the amplitude in the SHM requires $2000\alpha^2$ events. The Sgr+SHM requires more events to be observed than the SHM because the modulation arising from the Sgr stream and the SHM are out of phase, leading to a reduction in the modulation amplitude as shown in \Fig{fig:sgr1}. For the DD+SHM, we consider $\vmin$ values between $130$ km/s and $150$ km/s (see \Fig{fig:dd1}). This region approximately coincides with a $1$ keV bin centered at about $4.5$ keV, for a $50$ GeV DM particle scattering off xenon. The DD+SHM, for $\rho_\text{DD} / \rho_\text{SHM} = 2$, would require $170\alpha^2$ events to be detected at significance $\alpha$ sigma while the SHM would require roughly $7500\alpha^2$ events. The addition of the DD to the SHM significantly reduces the number of necessary events because the DD+SHM has a significantly larger modulation amplitude, as seen in \Fig{fig:dd1}. 

An important question to ask is, if an experiment were to view the annual modulation with a significance of $\alpha$ standard deviations, would additional events be necessary in order to distinguish these models? It is clear from \Fig{fig:sgr1} that the modulation features of the Sgr+SHM deviate most from those in the SHM in the phase of the first harmonic. Thus the question to ask for the Sgr+SHM is, how many events must be observed between $\vmin = 280$ km/s and $\vmin = 300$ km/s  in order to distinguish a phase occurring in mid March from a phase occurring in late May. The first harmonic in the DD+SHM differs strongly from that of the SHM in both the phase and the amplitude, and thus it is important to check which feature will be more easily distinguishable from the SHM.

We will begin with a rough analysis of the number of events necessary to discern the difference between the amplitude of the modulation for the DD+SHM, assuming $\rho_\text{DD} / \rho_\text{SHM} = 2$, from the SHM amplitude. Consider once again the same bin analysis previously used to determine the detectability of the modulation amplitude. The condition for distinguishing the modulation amplitude of the DD+SHM from the modulation amplitude of the SHM is simply that the difference between $\Delta N_\text{DD+SHM}$ and $\Delta N_\text{SHM}$ must be larger than the uncertainty, $\sqrt{N_\text{tot}}$, in the measurement of $\Delta N$. This implies 
\beq
\Delta N_\text{DD+SHM} - \Delta N_\text{SHM} \geqslant \alpha \sqrt{N_\text{tot}} \ .
\label{eq:ampdiscern1} 
\eeq
With similar manipulations as above, one can arrive at the following condition on $N_\text{tot}$: 
\beq
N_\text{tot} \geqslant \frac{\pi^2\alpha^2}{4} \left[\left(\frac{R_1}{R_0}\right)_\text{DD+SHM}-\left(\frac{R_1}{R_0}\right)_\text{SHM}\right]^{-2} \ .
\label{eq:ampdiscern2}
\eeq
Evaluating \Eq{eq:ampdiscern2} in the energy range previously defined for the DD+SHM, we find that approximately $225\alpha^2$ events are necessary to distinguish the amplitude of the DD+SHM from the SHM at a significance of $\alpha$ sigma. This implies that approximately $55\alpha^2$ more events must be detected after the annual modulation is observed in order to discriminate the DD+SHM from the SHM using only the amplitude of the modulation.

We now consider how the phase of the modulation could be used to estimate the number of events that must be observed in order to distinguish the various models. Assume for a moment that an annual modulation has been detected and the number of events can be plotted against time to form a sinusoidal-like figure. Let us assume that one of the data points lies at $(t_*,\bar{N})$, where $\bar{N}$ is the average number of events observed. Two cosine functions, one passing through the data point itself with phase $t_a$, and the other passing through the upper end of its error bar with phase $t_b$, can then be used to characterize the uncertainty with which the phase is known. The upper bound of the data point is proportional to the square root of the number of events in the temporal bin $\sqrt{N_\text{bin}}$. We will assume events are evenly distributed across temporal bins, implying $\sqrt{N_\text{bin}} \simeq \sqrt{N_\text{tot}}/4$. Assuming $\Delta N$ is known, the conditions that by definition must be satisfied are
\begin{align}
\frac{\Delta N}{2} \cos(\omega(t_*-t_a)) &= 0 \ , 
\label{eq:cosa}
\\  
\frac{\Delta N}{2} \cos(\omega(t_*-t_b)) &\simeq \frac{\alpha}{4} \sqrt{N_\text{tot}} \ .  \rule{0 mm}{10 mm}
\label{eq:cosb}
\end{align}
We solve \Eq{eq:cosa} for $t_*$ and restrict our attention to the solution that is closer in phase with the data. Substituting this result into \Eq{eq:cosb} yields 
\beq
\sin(\omega \Delta t) \simeq \frac{\alpha \sqrt{N_\text{tot}}}{2 \Delta N} \ ,
\label{eq:deltatime}
\eeq
where $\Delta t \equiv t_a - t_b$. To be conservative, we choose to restrict the uncertainty in the phase to be at most one month, for which $\sin(\omega \Delta t) = \sin(\pi/6) = 1/2$, which then implies
\beq
N_\text{tot} \geqslant \frac{\alpha^2 \pi^2}{4} \left( \frac{R_0}{R_1} \right)^2 \ ,
\label{eq:ntottime}
\eeq 
which coincidentally is the as same as \Eq{eq:ntot}. We note that the above analysis is only one sided in that it fails to account for the lower part of the $(t_*,\bar{N})$ error bar. The true uncertainty in the phase thus has a full width of two months, extending to one month to either side of the best-fit value. 

Since \Eq{eq:ntot} and \Eq{eq:ntottime} coincide, the number of events required to distinguish the phases (with a two month error) of the Sgr+SHM or DD+SHM modulations from the phase of the SHM modulation are approximately the same as those required to confirm the existence of the modulation itself. We thus expect any experiment measuring the modulation in an energy range where the phases of the models significantly differ, to measure the phase with high enough accuracy to differentiate the SHM from the Sgr+SHM and DD+SHM

\section{Conclusions}
\label{sec:conclusion}

We have considered how gravitational focusing of DM due to the Sun's gravitational potential would alter the time modulation of a DM signal. Previous studies have separately considered extracting information using a harmonic analysis \cite{LeeLisantiSafdi} and investigating how anisotropies in the DM halo might influence direct DM detection experiments \cite{AlenaziGondolo,LeeLisantiPeterSafdi,BozorgniaSchwetz}. The purpose of this paper is to unify these analyses and investigate how GF would alter the results of a harmonic analysis in the presence of DM velocity substructure.

We performed our analysis on a dark halo described by the standard halo model (SHM), a SHM with an added DM stream as expected from the tidal disruption of the Sgr dwarf galaxy by the Milky Way, a SHM plus a dark disk (DD) with lag speed $v_\text{lag}=50$ km/s, and a dissipative dark disk alone (DDD) with the same lag speed. Our results for the SHM alone are in agreement with Ref.~\cite{BozorgniaSchwetz}. Additionally, the conclusion of Ref.~\cite{LeeLisantiSafdi} that there should exist ratios of the amplitudes of harmonics independent of $\vmin$ was shown to be inconsistent with the presence of GF at $\vmin \lesssim 300$ km/s. This does not come as a surprise as the result of \cite{LeeLisantiSafdi} assumes that the local DM halo in the galactic frame is isotropic, and GF inherently makes the halo anisotropic.

For the Sgr stream, modeled with a velocity $\bol{v_\text{Sgr}}=(-65, 135, -249)$ km/s in galactic coordinates, we found that GF is unlikely to significantly affect any DM particles coming from the stream, but can affect the smooth halo component, and thus can alter the relative contributions of the Sgr stream and the smooth halo to the velocity integral $\eta(\vmin, t)$. We showed that by increasing the relative importance of the background halo, GF tends to reduce characteristic features that would otherwise be expected to appear in the phases of the annual and biannual harmonics from the inclusion of the Sgr stream component. In spite of this, GF does not eliminate the more prominent features which have the potential to alter the expected phase of the annual modulation by more than two months for values of $\vmin \approx v_\text{Sgr}^\text{S}$ with respect to the SHM alone, where $v_\text{Sgr}^\text{S}$ is the speed of the Sgr stream in the Sun's reference frame.

For our DD+SHM analysis we considered a dark disk co-rotating with the baryonic disk but with a smaller rotational velocity. Since the relative velocity of DM in the DD is much smaller than in the SHM or Sgr stream, one would expect GF to have a much larger influence in this model. Indeed we showed that the inclusion of a DD has a large influence on the unmodulated rate, the fractional amplitude, the amplitudes of the annual and biannual harmonics, and the phases of the annual and biannual harmonics. However, these effects appear only at $\vmin \lesssim 250$ km/s.

We also provided rough estimates of how many events should be observed in order to differentiate between the Sgr+SHM, DD+SHM, and SHM. We have determined that should an experiment measure the annual modulation in an energy range where the phase of the Sgr+SHM and DD+SHM differ noticeably from that of the SHM, the uncertainty in the measured phase will be small enough to allow for a discrimination between these models.

Our conclusions support the idea that analyzing the harmonic series of the DM differential scattering rate could potentially shed light on the distribution of DM in our galaxy. We have found that when DM velocity substructure is present, GF washes out some of the more distinctive features that would appear in the amplitudes and phases of the dominant harmonics were GF neglected. This is so because GF enhances the density of the low velocity WIMPs in the smooth halo component. However, deviations with respect to the SHM, most notably in the phases of the harmonics, can still persist and could provide insight into the astrophysical nature of DM.

\section{Acknowledgments}
We heartily thank Ji-Haeng Huh for his participation and support in the initial stages of this work. E.D.N.~and G.B.G.~were supported by the Department of Energy under Award Number DE-SC0009937.

\end{document}